\documentclass[11pt, a4paper, nofootinbib]{article}
\usepackage{graphicx}
\usepackage[bookmarks, colorlinks, breaklinks, pdfauthor={Huw Price}]{hyperref}
\usepackage{tikz}
\def\R2Lurl#1#2{\mbox{\href{#1}{\tt #2}}}

\usepackage{amsmath, amsthm, amssymb}

\usepackage{sectsty}
\subsubsectionfont{\hspace{-0.60cm}
\normalfont\emph}
\subsectionfont{\hspace{-0.60cm}\nohang\normalfont\normalsize\textit}
\sectionfont{\nohang\centering\normalfont\textsc}

\newcommand{\comment}[1]{}

\renewcommand{\marginpar}{\comment}

\hypersetup{linkcolor=blue,citecolor=blue,filecolor=black,urlcolor=blue}

\def\str#1{{\setbox1=\hbox{#1}\leavevmode 
\raise.47ex\rlap{\hspace{4.5pt}\leaders\hrule\hskip\wd 1} 
\box1}}

               \usepackage{latexsym}

\DeclareRobustCommand\textprime{\leavevmode \raise.8ex\hbox{\text@char\scriptfont\prime}}

\newcommand{\figia}{\begin{center}
\begin{tikzpicture}
\begin{scope}[scale=1.2]
\draw[color=black,ultra thin] (-2.5,-2)--(-2.5,2)--(2.5,2)--(2.5,-2)--(-2.5,-2); 
\draw[color=blue,densely dashed,->] (0.5,0)--(1,0);
\draw[color=blue,densely dashed,->] (0,0.5)--(0,1);
\draw[color=gray,ultra thick] (-0.5,-0.5)--(-0.5,0.5)--(0.5,0.5)--(0.5,-0.5)--(-0.5,-0.5)--(-0.5,0.5);
\draw[color=gray,thick] (-0.4,-0.4)--(0.4,0.4);
\draw (0,1.2) node {{\footnotesize $\textup{\textbf{Path R0}}$}};
\draw (1.6,0.03) node {{\footnotesize $\textup{\textbf{Path R1}}$}};
\draw[color=blue,thick, ->] (-1.9,0)--(-0.5,0); 
\draw (-1.25,-0.2) node {{\scriptsize $\tau$}};
\draw (-1.25,0.2) node {{\scriptsize Beam}};
\draw (0.15,-1) node {{\scriptsize\parbox{2cm}{Polarizing cube\\ set at angle $\sigma_R$}}};

\draw (0.,-2.5) node {{\small \parbox{8cm}{\begin{center}
\emph{Figure 1: A polarizing cube.}
\end{center}
}}};
\end{scope}
\end{tikzpicture}
\end{center}}

\newcommand{\figiia}{\begin{center}
\begin{tikzpicture}
\begin{scope}[scale=1.2]

\draw[color=black,ultra thin] (-2.5,-2)--(-2.5,2)--(2.5,2)--(2.5,-2)--(-2.5,-2); 

\draw[color=blue,densely dashed, <-] (-0.5,0)--(-1,0);
\draw[color=blue,densely dashed, <-] (0,0.5)--(0.0,1);
\draw[color=gray,ultra thick] (-0.5,-0.5)--(-0.5,0.5)--(0.5,0.5)--(0.5,-0.5)--(-0.5,-0.5)--(-0.5,0.5);
\draw[color=gray,thick] (-0.4,-0.4)--(0.4,0.4);
\draw (0,1.2) node {{\footnotesize $\textup{\textbf{Path L0}}$}};
\draw (-1.57,0.03) node {{\footnotesize $\textup{\textbf{Path L1}}$}};
\draw[color=blue,thick, ->] (0.5,0)--(1.9,0); 
\draw (1.25,-0.2) node {{\scriptsize $\tau$}};
\draw (1.25,0.2) node {{\scriptsize Beam}};
\draw (0.15,-1) node {{\scriptsize\parbox{2cm}{Polarizing cube\\ set at angle $\sigma_L$}}};

\draw (0.,-2.5) node {{\small \parbox{8cm}{\begin{center}
\emph{Figure 2: The polarizing cube in reverse.}
\end{center}
}}};
\end{scope}

\end{tikzpicture}
\end{center}}

\newcommand{\figiii}{\begin{center}
\begin{tikzpicture}
\begin{scope}[scale=1.2]

\draw[color=black,ultra thin] (-4.3,-2)--(-4.3,2)--(4.3,2)--(4.3,-2)--(-4.3,-2); 

\draw[color=blue,densely dashed] (2.5,0)--(3,0);
\draw[color=blue,densely dashed] (2,0.5)--(2.0,1);
\draw[color=gray,ultra thick] (1.5,-0.5)--(1.5,0.5)--(2.5,0.5)--(2.5,-0.5)--(1.5,-0.5)--(1.5,0.5);
\draw[color=gray,thick] (1.6,-0.4)--(2.4,0.4);
\draw (2,1.2) node {{\footnotesize $\textup{\textbf{Path R0}}$}};
\draw (3.6,0.03) node {{\footnotesize $\textup{\textbf{Path R1}}$}};
\draw (2,-1) node {{\scriptsize\parbox{2cm}{Polarizing cube\\ set at angle $\sigma_R$}}};

\draw[color=blue,densely dashed] (-2.5,0)--(-3,0);
\draw[color=blue,densely dashed] (-2,0.5)--(-2.0,1);
\draw[color=gray,ultra thick] (-1.5,-0.5)--(-1.5,0.5)--(-2.5,0.5)--(-2.5,-0.5)--(-1.5,-0.5)--(-1.5,0.5);
\draw[color=gray,thick] (-1.6,-0.4)--(-2.4,0.4);
\draw (-2,1.2) node {{\footnotesize $\textup{\textbf{Path L0}}$}};
\draw (-3.6,0.03) node {{\footnotesize $\textup{\textbf{Path L1}}$}};
\draw[color=blue,thick] (-1.5,0)--(1.5,0); 
\draw (0,-0.2) node {{\scriptsize $\tau$}};
\draw (0,0.2) node {{\scriptsize Photon}};
\draw (-2,-1) node {{\scriptsize\parbox{2cm}{Polarizing cube\\ set at angle $\sigma_L$}}};

\draw (0.,-2.5) node {{\small \parbox{8cm}{\begin{center}
\emph{Figure 3: Combining two cubes.}
\end{center}
}}};
\end{scope}

\end{tikzpicture}
\end{center}}

\newcommand{\figiv}{\begin{center}
\begin{tikzpicture}
\begin{scope}[scale=1.2]

\draw[color=black,ultra thin] (-4.3,-2)--(-4.3,2)--(4.3,2)--(4.3,-2)--(-4.3,-2); 

\draw[color=blue,densely dashed] (2.5,0)--(3,0);
\draw[color=blue,densely dashed] (2,0.5)--(2.0,1);
\draw[color=gray,ultra thick] (1.5,-0.5)--(1.5,0.5)--(2.5,0.5)--(2.5,-0.5)--(1.5,-0.5)--(1.5,0.5);
\draw[color=gray,thick] (1.6,-0.4)--(2.4,0.4);
\draw (2,1.2) node {{\footnotesize $\textup{\textbf{Path R0}}$}};
\draw (3.6,0.03) node {{\footnotesize $\textup{\textbf{Path R1}}$}};
\draw (2,-1) node {{\scriptsize\parbox{2cm}{Polarizing cube\\ set at angle $\sigma_R$}}};

\draw[color=blue,densely dashed] (-2.5,0)--(-3,0);
\draw[color=blue,densely dashed] (-2,0.5)--(-2.0,1);
\draw[color=gray,ultra thick] (-1.5,-0.5)--(-1.5,0.5)--(-2.5,0.5)--(-2.5,-0.5)--(-1.5,-0.5)--(-1.5,0.5);
\draw[color=gray,thick] (-1.6,-0.4)--(-2.4,0.4);
\draw (-2,1.2) node {{\footnotesize $\textup{\textbf{Path L0}}$}};
\draw (-3.6,0.03) node {{\footnotesize $\textup{\textbf{Path L1}}$}};
\draw[color=blue,thick] (-1.5,0)--(1.5,0); 
\draw (-1.2,-0.2) node {{\scriptsize $\tau_L$}};
\draw (1.15,-0.2) node {{\scriptsize $\tau_R$}};
\draw (0,-0.2) node {{\scriptsize ?}};
\draw (0,0.2) node {{\scriptsize Photon}};
\draw (-2,-1) node {{\scriptsize\parbox{2cm}{Polarizing cube\\ set at angle $\sigma_L$}}};

\draw (0.,-2.5) node {{\small \parbox{8cm}{\begin{center}
\emph{Figure 4: Time-symmetric polarization.}
\end{center}
}}};
\end{scope}

\end{tikzpicture}
\end{center}}

\begin{document}               
    \title{Dispelling the Quantum Spooks -- a Clue that Einstein Missed?}
\author{Huw Price\thanks{Trinity College, Cambridge CB2 1TQ, UK; email \href{mailto:hp331@cam.ac.uk}{hp331@cam.ac.uk}.} {\ and} Ken Wharton\thanks{Department of Physics and Astronomy, San Jos\'{e} State University, San Jos\'{e}, CA 95192-0106, USA; email \href{mailto:kenneth.wharton@sjsu.edu}{kenneth.wharton@sjsu.edu}.}}
\date{}           
\maketitle
\begin{abstract}\noindent It is well-known that Bell's Theorem and other No Hidden Variable theorems have a ``retrocausal loophole'', because  they assume that the values of pre-existing hidden variables are independent of future measurement settings. (This is often referred to, misleadingly, as the assumption of ``free will''.) However, it seems to have gone unnoticed until recently that a violation of this assumption is a straightforward consequence of time-symmetry, given an understanding of the quantization of light that would have seemed natural to Einstein after 1905. The new argument shows precisely why quantization makes a difference, and why time-symmetry alone does not imply retrocausality, in the classical context. It is true that later developments in quantum theory provide a way to avoid retrocausality, without violating time-symmetry; but this escape route relies on the ``ontic'' conception of the wave function that Einstein rejected. Had this new argument been noticed  much sooner, then, it seems likely that retrocausality would have been regarded as the default option for hidden variables theories (a fact that would then have seemed confirmed by Bell's Theorem and the No Hidden Variable theorems). This paper presents these ideas at a level intended to be accessible to general readers.
\end{abstract}\

\section{Einstein's dream}
Late in his life, Einstein told Max Born that he couldn't
take quantum mechanics seriously, ``because it cannot be
reconciled with the idea that physics should represent a reality in
time and space, free from spooky actions at a distance."\cite{spook47}
Most physicists think of this as the sad lament  of an old giant, beached on the wrong side of history -- unable to accept the revolution in physics  that he himself had done so much to foment, decades before, with his 1905 discovery that light is absorbed in discrete ``quanta''. 

According to this orthodox view, John
Bell put the final nail in the coffin of the dream of a spook-free successor to quantum theory, a decade after Einstein's death. Bell's Theorem seemed to show that any theory of the kind Einstein hoped for -- an addition to quantum theory, describing an underlying ``reality in space and time'' -- is bound to  involve the kind of ``action at a distance", or ``nonlocality'',  to which Einstein was objecting. 

So Einstein's spooks have long seemed well-entrenched in the quantum world -- and this despite the fact that they still seem to threaten  Einstein's other great discovery from 1905, special relativity.  As David Albert and 
Rivka Gal\-chen put it  in a recent
piece in \emph{Scientific American}, writing about the intuition of ``locality'': ``Quantum mechanics has upended many an intuition,
but none deeper than this one. And this particular upending carries
with it a threat, as yet unresolved, to special relativity---a
foundation stone of our 21st-century physics.''\cite{Albert}

But could this accepted wisdom be due for a shake-up? Could Einstein have the last laugh after all? Intriguingly, it turns out there's a new reason for taking seriously a  little-explored loophole in Bell's Theorem.\footnote{The same loophole exists in  the other  so-called ``No Hidden Variable theorems'', that seek to prove that there can't be a deeper reality underlying quantum theory, of the kind Einstein hoped for. For simplicity we'll  focus on Bell's Theorem, but the spook-dispelling technique we're talking about works equally well in the other cases.} Even more intriguingly, it's a reason that Einstein himself could have spotted as early as 1905, since it is a simple consequence of 
 the quantization of light, together with another assumption that he certainly accepted at that time.

The loophole stems from the fact that Bell's argument assumes that our measurement choices cannot influence the \emph{past} behaviour of the systems we choose to measure. This may seem quite uncontroversial. In the familiar world of our experience, after all, causation doesn't work ``backwards''. But a few physicists have challenged this assumption, proposing theories in which causation \emph{can} run backwards in the quantum world. This idea --
``retrocausality'', as it is often called -- has been enjoying a small
renaissance. (See Box 1.)

Until very recently,\cite{Price1}  however, no one seems to have noticed that there
is a simple argument that could have put retrocausality at centre-stage 
\emph{well before} the development of quantum theory. As we explain below, the argument shows that retrocausality follows directly from 
 the quantization of light, so long as fundamental physics is \emph{time-symmetric} (meaning that any physical process allowed in one time direction is equally allowed in reverse). 

Many ordinary physical processes, such as cream mixing in coffee, don't appear to be time-symmetric.  (The cream mixes in, but never mixes out!)  But these are processes involving large numbers of microscopic constituents, and the constituents seem to behave with complete time-symmetry. Cream ``unmixing'' out of coffee is physically possible, just very improbable.  Einstein himself defended the view that ``irreversibility depends exclusively upon reasons of probability'',\cite{Ritz} and is not evidence of a fundamental time-asymmetry.

\begin{figure}[t]
\input{box1.tex}
\end{figure}

As we shall explain, the new argument shows that quantization makes a crucial difference. Time-symmetry alone doesn't guarantee that causation ever works backwards, but quantization gives us a new kind of influence, which -- assuming time-symmetry  -- must work \emph{in both temporal
directions. } This new kind of influence is so subtle that it can evade spooky nonlocality,
without giving us an even more spooky ability to send signals into the
past. One of the striking things about the apparent action at a distance in quantum mechanics (QM) is that it, too, is subtle in just this way: there's no way to use it to build a ``Bell Telephone'', allowing superluminal communications. The argument hints how this subtlety might arise, as a consequence of quantization,  from an underlying reality that smoothly links everything together, via pathways permitted by relativity.

What would Einstein have thought, had he noticed this argument? We can only speculate, but one thing  is clear. If physics had already  noticed that quantization \emph{implies}
retrocausality (given fundamental time-symmetry, which was widely accepted), then it wouldn't have been possible for later physicists simply to ignore this option, as often happens today. 
It is true that the later development of quantum theory provides a way to avoid retrocausality, without violating time-asymmetry. But as we'll see,  this requires the very bit of quantum theory that Einstein disliked the most -- the strange, nonlocal wave function, that Einstein told Born he couldn't take seriously.

By Einstein's lights, then, the new argument seems important. If it had been on the table in the early years of quantum theory, the famous debates in the 1920s and 1930s about the meaning of the new theory couldn't possibly have ignored the hypothesis that a quantum world of the kind Einstein hoped for would need to be retrocausal.

Similarly, the significance of Bell's work in the 1960s would have seemed quite different.  Bell's work shows that \emph{if there's no retrocausality,} then QM is nonlocal, in apparent tension  with special relativity.\footnote{Bell himself certainly thought there was such a tension -- as he puts it at one point, ``the cheapest resolution [of the puzzle of nonlocality]
is something like going back to relativity as it was
before Einstein, when people like Lorentz and
Poincar\'e thought that there was an aether.''\cite{Bell1}}  Bell knew of the retrocausal loophole, of course, but was disinclined to explore it. (He once said that when he tried to think about backward causation he ``lapsed quickly into fatalism''.\cite{Bell2})
But if retrocausality had been on the table for half a century, 
 it would presumably have been natural for Bell and others to take this option more seriously.  Indeed, Bell's Theorem itself might then  have been interpreted as a \emph{second} strong clue that the quantum world is retrocausal -- the first clue (the one that Einstein missed) being the argument that the quantization of light implies retrocausality, given time-symmetry. 

In this counterfactual history, physicists in the 1960s --  two generations ago -- would already have known a remarkable fact about light quantization and special relativity, Einstein's two greatest discoveries from 1905. They would have known that both seem to point in the direction of retrocausality.\footnote{Quantization does so via the argument below. Relativity does so via Bell's Theorem, since the retrocausal loophole provides a way to escape the apparent tension between Bell's result and relativity.} What they would have made of this remarkable fact, we cannot yet say, but we think it's time we tried to find out.

\section{Polarization, From Classical to Quantum}

The new argument emerges most simply from thinking about a property of light called ``polarization''. (It can also be formulated using other kinds of properties, such as spin, but the polarization case is the one that Einstein could have used in 1905.) So let's start with the basics of polarization.
 
Imagine a wave travelling down a string.  Looking down the
string's axis you would see the string
moving in some perpendicular direction; horizontal, vertical, or at
some other angle.  This angle is strongly analogous to the
``polarization angle'' of an
electromagnetic wave in Classical Electromagnetism (CEM).  For a
linearly polarized CEM wave, the electric field is oscillating at some
polarization angle, perpendicular to the direction the wave is
travelling.

In a pair of polarized sunglasses are filters that allow light with one
polarization angle to pass through (vertical, 0$^\circ$),
while absorbing the other component (horizontal,
90$^\circ$).  For light with some other polarization angle
(say, 45$^\circ$), such a polarizing filter naturally
divides the electric field into horizontal and vertical components,
absorbing the former and passing the latter.  (The exact two components
depend on the angle of the filter, which you can confirm by watching
the blue sky gradually change color as you tip your head while wearing
polarized sunglasses.)

But there are also optical devices that \textit{reflect} one
polarization component while passing the other; an example of such a
``polarizing cube'' is depicted in Figure 1.  This
cube has a controllable angle $\sigma_R$ that splits the incoming light into two components.  (The portion polarized at angle  $\sigma_R$ passes
into the R1 path, and the portion polarized at
$\sigma_{R}+90^\circ$ reflects into the R0 path.\vspace{12pt}

\begin{figure}[h]
\figia
\end{figure}\vspace{-6pt}

A striking feature of CEM is that the equations are all
time-symmetric, meaning that if you can run any solution to the equations in reverse,
you'll have another solution.  Imagining the
time-reverse of Figure 1 is the easiest way to see that these
polarizing cubes can also combine two inputs into a single output beam,
as in Figure 2.  Different input combinations will produce different
output polarizations $\tau$. Putting all the
input beam on L1  will force $\tau$ to be equal to the
angle $\sigma_{L}$; an L0 input will force $\tau=\sigma_L+90^\circ$; and an appropriate mix of L1 and L0 can produce any polarization angle in between these extremes.   

\begin{figure}[t]
\figiia
\end{figure}

One of Einstein's great discoveries in 1905 -- he thought of it as the most important of all -- is that CEM is not always correct. It breaks down when the
energy in the electromagnetic field is very small --
i.e., at the levels of Einstein's light quanta, or photons, as we now call them. Einstein himself does not seem to have thought about the implications of this ``quantization'' for polarization, and the classic early discussion (to which Einstein later refers approvingly) is due to Paul Dirac \cite{Dirac}   in 1930. In terms of our example, the crucial point  is that in the case of a single photon, the light passing through cube in Figure 1 will \textit{not}
split into two paths.  Instead, when measured, the photon is always
found to be on R1 or on R0, not on both.  (The probabilities of these two
outcomes are weighted such that in the many-photon limit the CEM
prediction is recovered, on average.)

This change may seem innocuous, but it has profound implications.  At the
very least, it shows that CEM fails and has to change.  But how?  Einstein wrestled with the problem of how to combine quantization with the classical theory for a few years, but eventually moved on to what he thought of as the \emph{simpler} problem of general relativity! As for Dirac, he simply dodged the issue, by asserting that it wasn't the business of physics to look for a story about the underlying reality: ``Questions about what decides whether the photon [goes one way or the other] and how it changes its direction of polarization when it does [so] cannot be investigated by experiment and should be regarded as outside the domain of science.'' (\cite{Dirac}, p.~6)

To the extent that quantum theory does provide us with a picture,  the orthodox view is to let CEM run its normal course until the photon passes through the cube and encounters some detectors.\footnote{Technically, the orthodox view uses an equation that evolves a quantum state instead of electromagnetic fields, but the sentiment is the same.} Then, at that very last moment,   the measurement process 
``collapses'' all the light onto only one detector, either on R0 or R1, just as it arrives.

Einstein disliked this ``collapse'' -- it is what he had in mind when complaining about
``spooky action at a distance'' (since the R0 and R1 detectors may be far apart).  He thought that the so-called collapse was just the usual process of updating of our knowledge after acquiring new information, and didn't involve any real change in the physical system. And he hoped for a theory providing some ``hidden variables'', describing how the light quanta actually behave. On such a view, presumably, the photon will be regarded as ``making a choice'' as it passes through the cube -- though not necessarily a choice determined by any pre-existing property.\footnote{As Einstein says, ``Dirac \ldots\  rightly points out that it would probably be difficult, for example, to give a theoretical description of a photon such as would give enough information to enable one to decide whether it will pass [one way or the other through] a polarizer placed (obliquely) in its way.''\cite{EinDirac}} 

Neither Einstein nor Dirac seems to have considered the question of the time-symmetry of these simple polarization experiments. On the face of it, however, both a collapse picture and the ``choosy photon'' picture turn out to be incompatible with the principle that the world is time-symmetric at a fundamental level. To see what that means, and why it matters, let's turn to  time-symmetry.



\section{Time-Symmetry}

How can it be that the laws of fundamental physics are time-symmetric?  After all, as we already noted, many ordinary processes seem to have a clear temporal preference -- like cream mixing in coffee, they happen one way but not in reverse. But by the end of the nineteenth century, most physicists had come to the view that these cases are not fundamental. They reflect the statistical behaviour of huge numbers of microscopic constituents of ordinary stuff, and the fact that for some reason we still don't fully understand, stuff was much more organized -- had very low \emph{entropy,} as physicists say -- some time in the past.\footnote{See Sean Carroll's blog [\href{http://preposterousuniverse.com/eternitytohere/faq.html}{bit.ly/SCeternity}] for an accessible introduction.} But this explanation seems fully compatible with time-symmetry at the fundamental level. 

This didn't mean that fundamental physics \emph{has} to be time-symmetric. In principle, there might be some deep time-asymmetry, waiting to be discovered. But by this stage, a century ago, the tide was running firmly in the direction of fundamental symmetry. And certainly Einstein read it this way. Although he recognised that time-asymmetric fundamental laws were possible in principle, he thought them unlikely. Writing to Michele Besso in 1918, he says, ``Judging from all I know, I believe in the reversibility of elementary events.  All temporal bias seems to be based on `order'.''\cite{EinsteinBesso}

Since then, the evidence has changed in one important respect. We now know that there is a subtle time-asymmetry in particle physics. But there is a more general symmetry, called CPT-symmetry, that is widely regarded as fundamental (and is equivalent to time-symmetry for the cases we are discussing here). So for practical purposes, the consensus remains the same: time-symmetry (or strictly speaking CPT-symmetry) is something fundamental. 


\marginpar{{\scriptsize Some re-writing here. I think the original didn't stress the distinction between the reverse process being the \emph{same} and it being \emph{allowed} -- the latter is the one that matters.}}

Still, this attitude to time-symmetry ``in principle'' often co-exists with a distinctly time-\emph{asymmetric} working model  of the quantum world, as we can see by thinking about how CEM is typically modified to explain
Einstein's photons.  It turns out that both the ``collapse'' view and   the  ``choosy photon'' view described above are time-asymmetric.

To see why, we need to know that the crucial test for a time-symmetric theory is not that everything that happens should look exactly the same in reverse. For example, a rock travelling from Earth to Mars looks different in reverse -- it now goes from Mars to Earth -- even though the underlying mechanics  is thoroughly time-symmetric. 
 The test is that if the theory \emph{allows} one process then it also \emph{allows} the reverse process, so that we can't tell by looking at a video of the process whether it is being played forwards or backwards. (If one process isn't allowed, then it is easy to tell: if the video shows the illegal process, it must be being played backwards.)\vspace{24pt}

\figiii\vspace{6pt}

With this in mind,
 think of a video of the apparatus shown in Figure 3, for a case in which a photon came in from the L1 path and was then 
detected on the R1 path.\footnote{We can't literally make videos of individual photons, but we can make computer-generated videos showing what our theories say about photons, and they work just as well, for applying this test.} This video shows the photon entering through a cube set to $\sigma_{L}$, exiting through a cube set to $\sigma_{R}$, with a polarization $\tau=\sigma_{L}$ in between. 

Reversing the video shows a photon entering the apparatus through a cube set to $\sigma_{R}$, exiting through a cube set to $\sigma_{L}$, on the same paths as before, \emph{and still with a polarization} $\tau=\sigma_{L}$ \emph{in between.} And this isn't allowed by the theory (except in the special case where $\sigma_{L}=\sigma_{R}$), according to either of these models. On the contrary, if the photon enters through a cube set to $\sigma_{R}$, using Path 1, the intermediate polarization should be $\tau=\sigma_{R}$, not $\tau=\sigma_{L}$.
In these models, then, time-reversing a legal process often leads to an illegal process.


This time-asymmetry is entirely a result of quantization. In the classical case, the light can always be split between the two paths, at one end of the experiment or both. This ensures that the reversed version of any  process allowed by CEM in the apparatus in Figure 3 is also a legal process. In the classical case, therefore, we can't tell which way a video is being played.

Somehow, then, combining discrete inputs and outputs with CEM destroys time-symmetry (at least if one tries to preserve standard CEM between the two cubes, as  in the
collapse view and the  choosy
photon view).  In
the next section we'll explore why this
happened, and the surprising consequences of insisting that time-symmetry be saved.


\section{Control -- Classical and Quantum}

Does time-symmetry mean that the future controls the past just as much as the past controls the future? That can't be true in general, because CEM is perfectly time-symmetric, and yet it is easy to see that we can use the apparatus in Figure 3 to send signals (and hence, if we wish, control things) from past to future -- i.e., left to right -- but not in the other direction. If we feed in a classical light beam on path L1, then by controlling the angle $\sigma_{L}$ we can control $\tau$, and hence encode a signal in the light beam. By measuring $\tau$ on the right hand side, we can then decode the message, or control some piece of machinery. But we can't do that from right to left, of course, even though the equations of CEM are perfectly time-symmetric. So where does the difference come from?

The answer is that there's still a big difference between the left and the right of the experiment, the past and the future, despite the fact that the laws of CEM are time-symmetric. On the left side, we normally control whether the light comes from L0 or L1, and as we'll see in a moment, this is crucial to our ability to use the device to signal to the future. On the right side, we don't control how much of the light goes out via R0 and R1, and it's really this difference that explains why we can't signal to the past. 


We can check this diagnosis by imagining a version of the full experiment in which we \emph{don't} control the inputs on the left, so that our control, or lack of it, is symmetric on the two sides.  To help us imagine this, let's introduce a Demon (from the distinguished family whose ancestors have been so helpful at other points in the history of physics!) Let's give the Demon exactly the control on the left that Nature has (and we lack) on the right.  The Demon therefore controls the light beam intensities on the two channels L0 and L1. It knows what setting $\sigma_{L}$ we have chosen for the angle of the left cube, and shares with us the goal of having a single light beam emerge from the cube in the direction shown. The question is, who now controls the polarization $\tau$? Is it us, or the Demon, or is the control shared in some way?

It turns out that the Demon now has full control. The Demon can make  $\tau$ take any value, by choosing appropriate intensities for the two input beams on L0 and L1. So the Demon could use the device to signal to the future, by controlling the polarization -- but we can't! (See Figure 5, in Box 2, for a simple analogy -- we are in the same position as Kirk, and the Demon is Scotty, with full control of the direction of the ship, so long as she can use two thrusters.)

We can now see why in the ordinary case, we can't use the device in Figure 3 to signal to the past, \emph{even though the underlying physics is time-symmetric.} The problem is simply that we don't control the outputs, the amount of light that leaves the apparatus on the two beams R0 and R1. Nature controls that, not us. If we make a change in $\sigma_{R}$, Nature makes a corresponding change in the amount of light on the two beams R0 and R1, and nothing changes in the past. 

But now let's see what happens if we ``quantize'' our Demon, by insisting that it input a light beam on a single channel -- either L0 or L1, \emph{but not both at the same time.} It turns out that this makes a big difference to what we can control on the left side of the apparatus -- though still not quite enough to enable us to signal to the future!

With this new restriction on the Demon, choosing $\sigma_{L}$ for the orientation of the left cube means that there are only two possibilities for $\tau$. Either $\tau=\sigma_{L}$, in the case in which the Demon uses channel L1, or $\tau=\sigma_{L}+\textup{90}^{\circ}$,  in the case in which the Demon uses channel L0. This means that while we don't control $\tau$ completely, we do have a lot more control than before we quantized our Demon. We can now limit $\tau$ to one of just two possibilities, 90$^{\circ}$ different from one another. (See Box 2 again -- this is like the case in which Scotty can only use one thruster, giving Kirk some control over where the \emph{Enterprise} goes.)

Summing up,  the quantization restriction on the Demon gives us \emph{some} control from left to right, just by controlling the angle $\sigma_{L}$, of a kind we don't have without quantization. But now symmetry shows that if Nature is quantized in the same sense -- if Nature has to put the output beam on R0 or R1 \emph{but not both} -- then we have the same new kind of control from right to left, just by controlling the angle $\sigma_{R}$. If the underlying physics is time-symmetric, and if the outputs on the right are quantized, controlling  $\sigma_{R}$ gives us some control over the polarization of the photon, immediately \emph{before} it reaches the right hand cube!

\newpage\begin{center}
\textbf{Box 2: How discreteness makes a difference.}\\\vspace{12pt}

\fbox{\parbox{11.6cm}{
\begin{center}
\includegraphics[height=6cm]{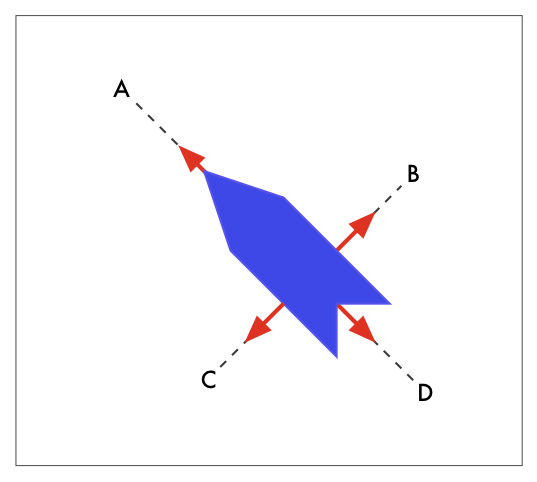}\\
\emph{Figure 5}
\end{center}
If Scotty has variable control of forward--reverse and left--right thrusters (shown as red arrows), then it doesn't matter which way Kirk points the \emph{Enterprise} (e.g., towards A, in Figure 5) -- Scotty can still make the ship travel in whatever direction she chooses, by putting an appropriate portion of the power on two different thrusters at 90$^{\circ}$ to each other. But if Scotty can only use \emph{one} thruster, then the \emph{Enterprise} has to travel \emph{either} in the orientation that Kirk chooses (along the line AD, in one direction or other), \emph{or} at 90$^{\circ}$ to that direction (along the line BC, in one direction or other). So Kirk now has at least \emph{some} control of the ship's direction, rather than none at all! This is exactly the difference that quantization makes, in the case of polarization. (For precision, we specify (i) that this \emph{Enterprise} lives in Flatland, and hence travels in only two dimensions, and (ii) that Scotty cannot vary the power of the thrusters during use -- any thruster in use burns for the same time, at its pre-set intensity.)}}
\end{center}\vspace{6pt}

To visualise this possibility, we need to modify our diagram a little. In place of the original $\tau$, controlled by $\sigma_{L}$, let's now write $\tau_{L}$. And let's introduce a new label, $\tau_{R}$, for the property that must be controlled by $\sigma_{R}$, if time-symmetry is to be preserved. This gives us Figure 4 -- we place $\tau_{L}$ and $\tau_{R}$ at left and right, respectively, next to the piece of the apparatus by whose setting they are (partially) controlled. (We'll come back to the question what might happen in the middle -- for the moment the ``?'' stands for ``unknown physics''!)

\figiv%

Summing up, we saw that  the quantization restriction on the Demon gave us some control over  $\tau_{L}$ by manipulating $\sigma_{L}$. Time-symmetry implies that an analogous  quantization restriction on Nature gives us control over  $\tau_{R}$ by manipulating $\sigma_{R}$. Again, we can't control $\tau_{R}$ completely, but we can limit it to two possibilities:  either $\tau_{R}=\sigma_{R}$ or $\tau_{R}=\sigma_{R}+\textup{90}^{\circ}$. (Note that this argument doesn't depend on the details of QM in any way -- we would have this new kind of control in CEM, if Nature cooperated in this way.)

\subsection{In Einstein's shoes}

Now think about this argument from Einstein's perspective, in 1905, after he's discovered that light is quantized, and before he's heard about quantum mechanics -- it lies twenty years in the future, after all! He knows that there's a minimum possible amount of light that can pass through the apparatus, what we now call a single photon. 

For a single photon, it seems, the quantization restriction must hold. The photon must travel on one path or other, \emph{at both ends of the experiment.} And that means that it is subject to this new kind of control -- the control that results from quantization -- \emph{in both directions.} (We are relying on time-symmetry here, of course -- if the laws governing the photon are allowed to be time-asymmetric, one sort of control, left to right, or right to left, can vanish.)

This simple argument shows that if we accept assumptions that would have seemed natural to Einstein after 1905, we do have a subtle kind of influence over the past, when we choose such things as the orientation of a polarizing cube. This is precisely the possibility called ``retrocausality'' in contemporary discussions of QM. As we noted above, most physicists think that they are justified in  ignoring it, and using ``No retrocausality'' as an uncontroversial assumption in various arguments intended to put further nails in the coffin of Einstein's view of the quantum world. And yet it turns out to have been hiding just out of sight, all these years!

\subsection{Spooks against retrocausality}

Defenders of orthodox QM, the view that Einstein came to dislike, are not going to be impressed by this argument. They'll point out that quantum theory allows us to avoid one or both of the two crucial assumptions, of time-symmetry and discrete (or quantized) outcomes. Both escape routes depend on the quantum wave function -- the strange object whose properties Einstein disliked so much. 

On some views, the wave function undergoes time-asymmetric ``collapse'', and so the underlying physics is not regarded as time-asymmetric. On other views, the wave function itself divides between the two output channels, R0 and R1. By varying the ``proportion'' of the wave function on each channel, Nature then has the flexibility she needs to respond to changes we make the angle  $\sigma_{R}$, preventing us from controlling the past. Either way, these versions of QM can escape the argument, and avoid retrocausality. 

But this escape from retrocausality rides on the back of the wave function -- the spooky, nonlocal object, not respectably resident in time and space, that Einstein deplored in his letter to Born. From Einstein's point of view, then, the argument seems to present us with a choice between spooks and retrocausality. We cannot say for certain which option he would have chosen -- ``There is nothing one would not consider when one is in a predicament!'', as he put it at one point, writing about the problem of reconciling CEM with quantization  \cite{Einstein1}  -- but the choice makes it clear that anyone who simply \emph{assumes} ``No retrocausality'' in arguing with Einstein is tying their famous opponent's arms behind his back. In the light of the argument above, retrocausality is the natural way of avoiding the spooks. To rule it out before the game starts is to deprive Einstein of what may be his star player.

\subsection{Control without signalling}
At this point, Einstein's contemporary opponents are likely to object that retrocausality is even worse than the so-called quantum spooks. After all, doesn't it lead to the famous paradoxes? Couldn't we use it to advise our young grandmother to avoid her unhappy marriage to grandfather, for example, and hence ensure that we had never been born? 

This is where the subtle nature of the new kind of control allowed by quantization is important. Let's go back to the case of the quantized Demon, who controls the inputs on the left  side of Figure 4, but can only use one channel at a time: all the input must be on either L0 or L1, and not on both at the same time. We saw that we do have some control over the polarization $\tau_{L}$ in this case, but not complete control. The Demon always has  the option of varying the polarization we aim for by 90$^{\circ}$. But this means that we can't control the polarization enough to send a signal -- intuitively, whatever signal we try to send, the Demon always has the option of turning it into exactly the opposite signal, by adding a factor of 90$^{\circ}$.\footnote{More formally, this is a consequence of the so-called No Signalling Theorem. A protocol that allowed us to signal in this case would also allow signalling between two arms of an analogous entanglement experiment with polarizers, in violation of the No Signalling Tteorem -- for the correlations in the two cases are exactly the same.\cite{EPW}}

So the kind of control introduced by quantization is too subtle to allow us to signal left to right, in the experiment in Figure 4. Hence, by symmetry, it is also too subtle to allow us to signal from right to left, or future to past. We couldn't use it to send a message to our grandmother, and history seems safe from paradox. Despite this, we can still have some control over hidden variables in the past, of the kind needed for retrocausality to resolve the challenge of Bell's Theorem in Einstein's favor.

\section{Where next?}

So discussions of the meaning of quantum theory  might have been very different, if Einstein had noticed how quantization suggests retrocausality. But would it have made any difference in the end? Is there any practical way to rescue Einstein's dream of a spook-free quantum world, if we do allow it to be retrocausal?

As we noted earlier (see Box 1), there are several retrocausal proposals on the table. But some, like the Aharonov-Vaidman Two State proposal, or the earlier Transactional Interpretation, try to build their retrocausal models with the same kind of elements that Einstein objected to -- wave functions not properly located in space and time.

If we want to stay close to the spirit of Einstein's program, then, we'll need to start somewhere else. One such proposal aims to extract a realistic model of quantum reality from the techniques developed by another of the twentieth century's quantum giants, Richard Feynman. Feynman's ``path integral''  computes quantum probabilities by considering all the possible ways a quantum system might get from one event to another, and assigning a weight, or ``amplitude'' to each possible ``path''.\footnote{These ``paths''  are not necessarily localized trajectories of particles, but could be the entire  ``histories'', between the two events in question, of something spread out in space, such as an electromagnetic field. (This extension from the particle-path integral to a field-history integral is used in quantum field theory.)}

The path integral is usually regarded simply as a calculational device. Most people, including Feynman himself, have thought that it doesn't make sense to interpret it as telling us that the system \emph{really} follows one particular history, even if we don't know which. Attractive as this thought seems in principle, the probabilities of the various paths just don't add up in the right way. 
But there are some clues that this might be made to work, using non-standard versions of the path integral.\footnote{As in a few recent efforts \cite{PathInt}\cite{Kent}; see also \cite{Sork}.}

Applied to the case we have been discussing, this approach suggests  that electromagnetic fields are not strictly constrained by the equations of CEM (a possibility that Einstein certainly contemplated).  Instead, the path integral considers \emph{all} possible field histories, even apparently crazy cases where a photon's polarization \emph{rotates} from $\tau_L$ to $\tau_R$ in empty space.  In fact, only by considering such non-classical cases can this approach make accurate predictions.  Such a scenario, involving a real polarization rotation between the cubes, is exactly the sort of process that could restore time-symmetry to the case of single photons.

Any approach that does succeed in interpreting the Feynman path integral realistically  -- i.e., that makes sense of the idea that the system actually follows just one of the possible histories that make up the path integral -- is likely to be retrocausal. Why? Simply because we have control over some aspects of the ``endpoint'' any path, when we choose to make a particular measurement (e.g., when we choose $\sigma_R$ in Figure 4). In general, our choice makes a difference to the possible paths that lead to this endpoint, and if one of those paths must be the real path followed by the photon, then we are affecting that reality, to some extent.

In putting future and past on an equal footing, this kind of approach is different in spirit from (and quite possibly formally incompatible with) a more familiar style of physics: one in which the past continually generates the future, like a computer running through the steps in an algorithm. However, our usual preference for the computer-like model may simply reflect an anthropocentric bias. It is a good model for creatures like us, who acquire knowledge sequentially, past to future, and hence find it useful to update their predictions in the same way. But there is no guarantee that the principles on which the universe is constructed are of the sort that happens to be useful to creatures in our particular situation.

Physics has certainly overcome such biases before -- the Earth isn't the center of the universe, our
sun is just one of many, there is no preferred frame
of reference. Now, perhaps there's one further anthropocentric
attitude that needs to go: the idea that the universe is
as ``in the dark'' about the future as we are ourselves.

It is too early to be sure that the spooks are going to be driven out of the quantum world; too early to be entirely confident that retrocausality will rescue Einstein's dream of a rela\-tivity-friendly quantum reality, living in space and time. What is clear, we think, is that there are some intriguing hints that point in that direction. Most intriguing of all, there's a simple new argument that suggests that  reality \emph{must} be retrocausal, by Einstein's lights, if we don't go out of our way to avoid it. It is too early to award Einstein the last laugh, certainly;  but too soon to dismiss the dreams of the old giant.

\end{document}